\documentstyle[12pt,a4,epsf]{article}
\begin{document}
\setlength{\textheight}{7.4in}
\setlength{\topmargin}{-0.6in}
\setlength{\oddsidemargin}{0.2in}
\setlength{\evensidemargin}{0.5in}
\renewcommand{\thefootnote}{\fnsymbol{footnote}}
%\centerline{\Large{\bf Draft only ! Please correct where necessary !}} 
\begin{center}
{\Large {\bf Instanton-Sphaleron Transition in the $d = 2$
Abelian--Higgs Model on a Circle}}

\vspace{0.8cm}

{D.K. Park,\raisebox{0.8ex}{\small a,b}\footnote[1]
{ Email:dkpark@hep.kyungnam.ac.kr}
H.J.W. M\"uller--Kirsten,\raisebox{0.8ex}{\small a}\footnote[2]
{Email:mueller1@physik.uni-kl.de}
J. Q. Liang\raisebox{0.8ex}{\small a,c}\footnote[3]
{Email:jqliang@mail.sxu.edu.cn}
and A.V. Shurgaia\raisebox{0.8 ex}{\small a,d}\footnote[4]
{Email:avsh@rmi.acnet.ge}}

\raisebox{0.8 ex}{\small a)}{\it Department of Physics, University
of Kaiserslautern, 67653 Kaiserslautern, Germany}

\raisebox{0.8 ex}{\small b)}{\it  Department of Physics, Kyungnam University,
Masan, 631--701, Korea}

\raisebox{0.8 ex}{\small c)}{\it Department of Physics and
Institute for Theoretical Physics, Shanxi University, Taiyuan, Shanxi 030006,
 P.R. China}

\raisebox{0.8 ex}{\small d)}{\it Mathematical Institute, Georgian Academy
of Sciences, Ruchadze Str., 380093 Tbilisi, Georgia}
\end{center}

\vspace{0.3cm}

{\centerline{\bf Abstract}}
\noindent
The transition from the instanton--dominated quantum
regime to the sphaleron--dominated
classical regime is studied in the $d=2$ Abelian--Higgs
model when the spatial
coordinate is compactified to $S^1$. Contrary to the noncompactified 
case, this model allows both sharp first--order and smooth second--order
transitions depending on the  
size of the circle. This finding may make the
model a useful toy model for the analysis of baryon number violating
processes. Since the model can to a large extent be treated analytically,
it can also serve as a transparent prototype for the
application of our method to more complicated cases, such as those
in higher dimensions.

%\documentstyle[prl,preprint,aps,epsf]{revtex4}
%\documentstyle[prl,preprint,aps,epsf]{revtex}
%\documentstyle[12pt]{article}
%\usepackage{amsmath}
%\usepackage{latexsym}
%\begin{document}

%\title{Instanton-Sphaleron transition in the d = 2
%Abelian-Higgs model on a Circle}

%\author{D.K. Park$^{1,2}$\footnote{e-mail:
%dkpark@hep.kyungnam.ac.kr},
%H. J. W. M\"{u}ller-Kirsten$^{1}$\footnote{e-mail:
%mueller1@physik.uni-kl.de}, J.--Q.
%Liang$^{1,3}$\footnote{e-mail:jqliang@mail.sxu.edu.cn} and
%A. V. Shurgaia $^{1}$\footnote{email: shurgaia@physik.uni--kl.de}}
%\address{1. Department of Physics, University of Kaiserslautern,
% D-67653 Kaiserslautern, Germany\\
%2.Department of Physics, Kyungnam University, Masan, 631-701, Korea\\
%3. Department of Physics and Institute of Theoretical Physics,
%Shanxi University, Taiyuan, Shanxi 030006, China}
%\maketitle
%\date{\today}
%\maketitle
%\begin{abstract}
%\vspace{8cm}

\vspace{2cm}

\section{Introduction}

After the sphaleron solution in the Weinberg--Salam model
had been found\cite{man83,kli84}, 
the temperature dependence of baryon number violating processes (BNVP) was
 studied extensively. To understand the overall features of BNVP
over the entire range of temperature, the computation of periodic 
instantons\cite{khl91} and their corresponding classical actions is required. 
However, the calculation of these in the Weinberg--Salam model is
a highly 
non--trivial problem, even if numerical techniques are employed. Hence 
in many cases simple toy models were used to explore the temperature 
dependence of BNVP.

An immediate candidate as a simple toy model is the $d=2$ Mottola--Wipf(MW) 
model\cite{mot89}, which shares many common features with $d=4$ 
electroweak theory. The scale invariance of the nonlinear $O(3)$ model is 
broken in the MW model by adding an explicit mass term. This has a close
analogy to the fact that the conformal invariance of the electroweak
theory is broken in the Higgs sector. Also, neither model supports 
a vacuum instanton which gives a dominant contribution to the winding 
number transition at low temperature. The transition between thermally
assisted quantum tunneling dominated by periodic instantons
 and the classical
crossover dominated by the sphaleron in the MW model has been analyzed in 
Refs.\cite{hab96,park99-1} using the method of
\cite{rana}, and it has been
 shown that the instanton--sphaleron
transition is of the sharp first--order type
 in the full range of parameter space.

Recently, however, a numerical study\cite{bon99,fro99} of 
the $d=4$ SU(2)--Higgs model -- which is 
a bosonic sector of the electroweak theory -- has
 shown that a smooth 
second--order transition occurs when $6.665 < M_H / M_W < 12.03$ although
the first--order transition occurs when $M_H / M_W < 6.665$\cite{higgs}.
 This implies that the MW model
does not exhibit a proper transition of BNVP when heavy Higgs's are involved.

Another candidate as a toy model is the $d=2$ Abelian-Higgs
 model which supports
vortex solutions\cite{niel73},
in particular the vacuum instanton and the sphaleron\cite{boch87}
simultaneously. The simultaneous existence of instanton and sphaleron 
causes the model to yield phase diagrams 
for the instanton--sphaleron
transition which are completely
different from those of electroweak theory, as shown in Ref.\cite{kuz97}. 
Furthermore, numerical\cite{mat93} and analytical\cite{park00-1} approaches
have shown that the instanton--sphaleron transition in this model is 
always of the second--order type,
 regardless of the ratio $M_H / M_W$. Hence, contrary
to the MW model, the ordinary Abelian--Higgs model does not describe the 
instanton--sphaleron transition of the electroweak theory
properly when the Higgs mass is 
small.

In the following 
we study the instanton--sphaleron transition in the $d=2$ 
Abelian-Higgs model when the spatial coordinate is compactified to 
$S^1$.
Quite apart from the question of the
physical relevance of the investigation below, it is a natural theoretical
curiosity to inquire what the order of
thermal transitions would be in this case, and
we present the answer here.
Physically, of course, the transitions we investigate are
not those with respect to an order parameter as in the
Weinberg--Salam theory, but with respect to temperature
or inverse period of the periodic instantons in the potential barrier. 
These transitions have physically the meaning of
transitions between classical and quantum behavior \cite{ourprl}.
Nonetheless, as stated we consider the model as an analogy,
which enables us to investigate corresponding behavior. 
Since, to our knowledge, the effect of the compactification 
of the spatial coordinate of this model has not yet
been investigated, this is also 
of interest on its own. Furthermore, we show that this model 
exhibits both first--order and second--order transitions
 depending on the
size of the circumference of the spatial coordinate domain,
i.e. the first order transition disappears in the limit
of the circumference becoming infinitely large, in fact, even beyond a
finite critical value. 
This means that the
Abelian--Higgs model defined on a circle can be a better toy model than
the MW model or the uncompactified Abelian-Higgs model for
an analysis which can be compared with  
that of BNVP. 
One may wonder how, if at all, this situation compares with finite
 size scaling
effects, i.e. of lattices with periodic boundary conditions, in lattice
gauge theory contexts. In the latter, see e.g.\cite{lattice}, lattice sizes
of $4^4$ to $16^{16}$ are used
and the possible dependence of the order of thermal transitions on
these is investigated. Nonetheless the lattice
sizes used are
presumably still too small to
permit definite conclusions about
the scaling regime.  We do not think that our case is really comparable to that.
Rather we view the present model as a testing ground for various aspects
related to phase transitions, since the study of the latter, as can be seen from the computations needed
in the following, are highly nontrivial
(and we therefore have to present some technical details),
 so that any model that can be handled
to a large extent analytically, is worth
 studying. Thus before higher dimensional
cases can be attacked convincingly,
 it is essential to have a thorough understanding
of a lower dimensional one
like the one we study here. 
This is therefore another main objective of the following.

\section{The Sphaleron Configuration}

We begin with the Euclidean action
\begin{equation}
\label{action}
S_E^{(0)} = \int d\tau dx 
      \left[ \frac{1}{4} F_{\mu \nu} F_{\mu \nu} + (D_{\mu} \phi)^{\ast}
             D_{\mu} \phi + \lambda [\mid \phi \mid^2 - \frac{v^2}{2}]^2
                                                            \right]
\end{equation}
and its field equations
\begin{eqnarray}
\label{motion}
\partial_{\mu} F_{\mu \nu}&=& ig 
\left[ \phi^{\ast} (D_{\nu} \phi) - (D_{\nu} \phi)^{\ast} \phi \right],
                                                          \\ \nonumber
D_{\mu} D_{\mu} \phi&=& 2 \lambda \phi (\mid \phi \mid^2 - \frac{v^2}{2}), 
\end{eqnarray}
where $D_{\mu} = \partial_{\mu} - i g A_{\mu}$.
We define as mass--dimensional parameters
\begin{eqnarray}
\label{mass}
M_H&\equiv&\sqrt{2 \lambda} v,   \\  \nonumber
M_W&\equiv&g v,
\end{eqnarray}
which correspond to Higgs mass and gauge particle mass
in electroweak theory respectively.
It is easy to show that the static sphaleron solution
in the $A_0=0$ gauge
is given by
\begin{eqnarray}
\label{sphaleron1}
A_1&=&A = const,    \\   \nonumber
\phi_{sph}&=&\frac{k b(k)}{\sqrt{\lambda}} e^{igAx} sn[b(k) x],
\end{eqnarray}
where $sn[z]$ is a Jacobian elliptic function, 
$k$ is the modulus of the elliptic function, and
\begin{equation}
\label{b(k)}
b(k) = \sqrt{\frac{\lambda}{2}} v 
      \left( \frac{2}{1 + k^2} \right)^{\frac{1}{2}}.
\end{equation}
Since $sn[z]$ has period $4K(k)$, where $K(k)$ is the complete elliptic
integral of the first kind, the circumference $L$ of $S^1$ is defined by
\begin{equation}
\label{circumference}
L_n = \frac{4 n K(k)}{b(k)},
\hspace{2.0cm} n = 1, 2, 3 \cdots.
\end{equation}
Since the transition rate is negligible for large $n$\cite{soo99}, 
we restrict ourselves to the $L = L_1$ case here.
In view of $K(1) =\infty$ we see that $k=1$ gives the uncompactified
limit we investigated previously\cite{park00-1}. Thus, since
this case does not lead to a first order transition, we can
expect one, if at all, only in the domain of small values of 
the elliptic modulus $k$, and in fact, we shall see that this is the case.

In order to examine the type of instanton--sphaleron transition we 
have to introduce the fluctuation fields around the sphaleron and 
expand the field equations (\ref{motion}) up to the third order in these 
fields. If, however, one expands Eq.(\ref{motion}) naively, one will
realize that the fluctuation operators are not diagonalized and, hence, 
the computation of the spectra of these operators becomes a very
non--trivial problem. To avoid this difficulty, we choose the 
$R_{\xi}$ gauge\cite{car90} by adding to the original
action (\ref{action}) the gauge fixing term
\begin{equation}
\label{fixing}
S_{gf} = \frac{1}{2 \xi} \int d\tau dx
\left[ \partial_{\mu} A_{\mu} + \frac{ig}{2} \xi (\phi^2 - \phi^{\ast 2})
                                                           \right]^2.
\end{equation}
 Then the field equations for the
total Euclidean action $S_E = S_E^{(0)} + S_{gf}$ become
\begin{eqnarray}
\label{eqmotion}
\partial_{\mu} F_{\mu \nu} + \frac{1}{\xi}
\left[ \partial_{\mu} \partial_{\nu} A_{\mu} + ig\xi (\phi \partial_{\nu}
       \phi - \phi^{\ast} \partial_{\nu} \phi^{\ast}) \right]
&=& ig \left[\phi^{\ast} (D_{\nu} \phi) - (D_{\nu} \phi)^{\ast} \phi \right],
                                             \\  \nonumber
D_{\mu}D_{\mu} \phi + ig\phi^{\ast}
\left[\partial_{\mu}A_{\mu} + \frac{ig\xi}{2} (\phi^2 - \phi^{\ast 2})
                                                            \right]
&=& 2 \lambda \phi (\mid \phi \mid^2 - \frac{v^2}{2}).
\end{eqnarray}
One can show that the sphaleron in this gauge is
the same as that of 
Eq.(\ref{sphaleron1}) if $A=0$:
\begin{eqnarray}
\label{sphaleron2}
A_1&=&0 ,    \\   \nonumber
\phi_{sph}&=&\frac{k b(k)}{\sqrt{\lambda}} sn[b(k) x].
\end{eqnarray}
We have therefore determined the sphaleron configuration
in the most optimal way to permit continuation with
the following difficult computations.

\section{Fluctuations about the Sphaleron}

We now introduce the fluctuation fields around the sphaleron
configuration by setting 
\begin{eqnarray}
\label{fluctuation}
A_0(\tau, x)&=& a_0(\tau, x),    \\  \nonumber
A_1(\tau, x)&=& a_1(\tau, x),    \\  \nonumber
\phi(\tau, x)&=& \phi_{sph}(x) + \frac{1}{\sqrt{2}} \bigg(\eta_1(\tau, x)
+ i \eta_2(\tau, x) \bigg),
\end{eqnarray}
where $a_0$, $a_1$, $\eta_1$, and $\eta_2$ are real fields. Inserting
(\ref{fluctuation}) into Eq.(\ref{action}) and Eq.(\ref{fixing}) 
one can express $S_E$ for $\xi=1$ as  
\begin{equation}
\label{expaction}
S_E = \frac{E_{sph}}{T} + S_2 + S_3 + S_4
\end{equation}
where $1/T$ is the period of the sphaleron \cite{note,yellow} and 
\begin{eqnarray}
E_{sph}&=& \sqrt{2 \lambda} v^3
          \bigg[\left[ \left( \frac{2}{1 + k^2} \right)^{-\frac{1}{2}}
                      + \frac{1 + 2 k^2}{3} \left(\frac{2}{1 + k^2}
                                            \right)^{\frac{3}{2}}
                      -2 \left(\frac{2}{1 + k^2} \right)^{\frac{1}{2}}
                                                            \right] K(k)
                                                         \\   \nonumber
& & 
\hspace{2.5cm}
               + \left[2 \left( \frac{2}{1 + k^2} \right)^{\frac{1}{2}}
                      - \frac{1 + k^2}{3} \left(\frac{2}{1 + k^2}
                                          \right)^{\frac{3}{2}}
                                                           \right] E(k)
                                                                \bigg],
                                                             \\  \nonumber
S_2&=& \int d\tau dx 
\Bigg[ \frac{1}{2} a_0 [-\partial_{\mu} \partial_{\mu} + 2 g^2 
                            \phi_{sph}^2] a_0 
      + \frac{1}{2} a_1 [-\partial_{\mu} \partial_{\mu} + 2 g^2 
                            \phi_{sph}^2] a_1
                                                  \\  \nonumber
& & \hspace{0.1cm}      
      + \frac{1}{2} \eta_1 \left[-\partial_{\mu} \partial_{\mu} 
                                 + 2 \lambda (3 \phi_{sph}^2 - \frac{v^2}{2})
                                                            \right] \eta_1
      + \frac{1}{2} \eta_2 \left[ -\partial_{\mu} \partial_{\mu} 
                                  + 2(\lambda + g^2) \phi_{sph}^2
                                     -\lambda v^2 \right] \eta_2
                                                              \\ \nonumber
& & \hspace{2.5cm}
      + 2 \sqrt{2} g \phi_{sph}^{\prime} a_1 \eta_2     \Bigg],
                                                              \\  \nonumber
S_3&=& \int d\tau dx
     \Bigg[ 2g (a_0 \dot{\eta_1} \eta_2 + a_1 \eta_1^{\prime} \eta_2)
            + \sqrt{2} g^2 \phi_{sph} (a_0^2 + a_1^2) \eta_1
                                                              \\  \nonumber
& & \hspace{2.0cm}
            + \sqrt{2} \lambda \phi_{sph} \eta_1^3 + 
              \sqrt{2} (\lambda + g^2) \phi_{sph} \eta_1 \eta_2^2
                                                                \Bigg],
                                                          \\   \nonumber
S_4&=& \int d\tau dx
      \left[ \frac{g^2}{2} (a_0^2 + a_1^2) (\eta_1^2 + \eta_2^2)
            + \frac{\lambda}{4} (\eta_1^2 + \eta_2^2)^2
            + \frac{g^2}{2} \eta_1^2 \eta_2^2        \right].
\end{eqnarray}
In these equations $E(k)$ is the complete elliptic integral
 of the second kind,
and the dot and the prime denote differentiation with respect to $\tau$ and
$x$ respectively. Owing to the final term in $S_2$ the fluctuation operators
for $a_1$ and $\eta_2$ are not diagonalized although the
$R_{\xi=1}$ gauge has been 
chosen. 
To obtain the diagonalization we introduce the fluctuation fields
$\rho_{\pm}$ defined as
\begin{equation}
\label{redefine}
\rho_+= v_1 a_1 + v_2 \eta_2 ,  \;\;
\rho_-= -v_2 a_1 + v_1 \eta_2,
\end{equation}
where 
\begin{equation}
\label{v1v2}
v_1=\sqrt{\frac{1 - (\phi_{sph}^2 - \frac{v^2}{2}) f_1^{-\frac{1}{2}}}{2}},
 \;\;                                                                
v_2=\sqrt{\frac{1 + (\phi_{sph}^2 - \frac{v^2}{2}) f_1^{-\frac{1}{2}}}{2}},
\end{equation}
and
\begin{equation}
\label{f1define}
f_1 = (\phi_{sph}^2 - \frac{v^2}{2})^2\cosh^2\alpha   - 
      \frac{v^4}{4} \left( \frac{1 - k^2}{1 + k^2} \right) \sinh^2 \alpha.
\end{equation}
Here $\alpha = \sinh^{-1} 2 \theta$ and $\theta$ is the
dimensionless parameter
\begin{equation}
\theta \equiv \frac{2 M_W}{M_H} = \sqrt{\frac{2 g^2}{\lambda}}.
\end{equation}
Using the field redefinition (\ref{redefine}) and the first--order differential
equation for $\phi_{sph}$, 
\begin{equation}
\label{firsteq}
\phi_{sph}^{\prime} + \sqrt{\lambda}
\left[ \frac{v^4}{4} \left( \frac{2k}{1 + k^2} \right)^2
      - v^2 \phi_{sph}^2 + \phi_{sph}^4  \right]^{\frac{1}{2}} = 0,
\end{equation}
it is straightforward to show that $S_2$ becomes
\begin{equation}
\label{s2dia}
S_2 = \frac{1}{2} \int d\tau dx
[a_0 D_0 a_0 + \eta_1 D_1 \eta_1 + \rho_+ D_+ \rho_+ + \rho_- D_- \rho_-],
\end{equation}
where
\begin{eqnarray}
\label{doperator}
D_0&=&-\partial_{\mu} \partial_{\mu} + 2 g^2 \phi_{sph}^2,
                                                     \\   \nonumber
D_1&=&-\partial_{\mu} \partial_{\mu} + 2 \lambda (3 \phi_{sph}^2 - 
                                                  \frac{v^2}{2} ),
                                                      \\   \nonumber
D_{\pm}&=&-\partial_{\mu} \partial_{\mu} + 2 g^2 \phi_{sph}^2 
          + \lambda (\phi_{sph}^2 - \frac{v^2}{2}) \mp \lambda 
                                                        \sqrt{f_1}.
\end{eqnarray}
After inserting the field redefinition (\ref{redefine}) into $S_3$ and
$S_4$, one can derive the field equations for the fluctuation fields
by varying the total action $S_E$, i.e.
(the method of Ref. \cite{rana} to determine the order of
thermal transitions requires all the terms written out explicitly
here)
\begin{equation}
\label{expand}
\hat{l} \left( \begin{array}{c}
               a_0 \\ \rho_+ \\ \rho_- \\ \eta_1
               \end{array}                      \right)
= \hat{h}
        \left(  \begin{array}{c}
                a_0 \\ \rho_+ \\ \rho_- \\ \eta_1
                \end{array}                      \right)
+       \left(  \begin{array}{c}
                G_2^{a_0} \\ G_2^{\rho_+} \\ G_2^{\rho_-} \\ G_2^{\eta_1}
                \end{array}                       \right)
+       \left(  \begin{array}{c}
                G_3^{a_0} \\ G_3^{\rho_+} \\ G_3^{\rho_-} \\ G_3^{\eta_1}
                \end{array}                       \right)
+ \cdots 
\end{equation}
where
%\begin{eqnarray}
%\label{explain}
$$
\begin{array}{lc}
{\hat{l}=\left( \begin{array}{clcr}
                 \frac{\partial^2}{\partial z_0^2} & 0 & 0 & 0 \\

                 0 & \frac{\partial^2}{\partial z_0^2} & 0 & 0 \\
 
                 0 & 0 & \frac{\partial^2}{\partial z_0^2} & 0 \\
 
                 0 & 0 & 0 & \frac{\partial^2}{\partial z_0^2} \\

                 \end{array}   \right),} &             
%\nonumber

%\end{eqnarray}
%and
%\begin{eqnarray}
%\label{see}
%\hspace{0.5cm}
{\hat{h} = \left( \begin{array}{clcr}
                 \hat{h}_{a_0} & 0 & 0 & 0  \\
                 0 & \hat{h}_{\rho_+} & 0 & 0  \\
                 0 & 0 & \hat{h}_{\rho_-} & 0  \\
                 0 & 0 & 0 & \hat{h}_{\eta_1} \\ 
                 \end{array}                     \right),}
\end{array}   
%\label{explain}
%\end{equation}
$$
and
\begin{eqnarray}
\label{explain}
G_2^{a_0}&=& \frac{2g}{b(k)} (v_2 \rho_+ + v_1 \rho_-) \dot{\eta_1}
            + \frac{2\sqrt{2}g^2}{b^2(k)} \phi_{sph} a_0 \eta_1,
                                                           \\ \nonumber
G_3^{a_0}&=& \frac{g^2}{b^2(k)} a_0
             \left[\eta_1^2 + (v_2 \rho_+ + v_1 \rho_-)^2 \right],
                                                             \\ \nonumber
G_2^{\rho_+}&=& \frac{2g}{b(k)}
                [v_2 a_0 \dot{\eta_1} + (v_1^2 - v_2^2) \rho_- \eta_1^{\prime}
                 + 2 v_1 v_2 \rho_+ \eta_1^{\prime}],
                                                             \\ \nonumber
& &+ \frac{2\sqrt{2} \lambda}{b^2(k)} \phi_{sph}
     [v_2^2 \rho_+ \eta_1 + v_1 v_2 \rho_- \eta_1]
   + \frac{2\sqrt{2} g^2}{b^2(k)} \phi_{sph} \rho_+ \eta_1,
                                                             \\ \nonumber
G_3^{\rho_+}&=& \frac{g^2}{b^2(k)}
               \Bigg[\rho_+ \eta_1^2 + v_2^2 a_0^2 \rho_+ 
                     + v_1 v_2 a_0^2 \rho_- + 2 v_1^2 v_2^2 \rho_+^3
                     + 3 v_1 v_2 (v_1^2 - v_2^2) \rho_+^2 \rho_-
                                                            \\ \nonumber
& & \hspace{2.0cm}
                    + (v_1^4 - 4 v_1^2 v_2^2 + v_2^4) \rho_+ \rho_-^2
                    - v_1 v_2 (v_1^2 - v_2^2) \rho_-^3  \Bigg] 
                                                              \\ \nonumber
& + & \frac{\lambda}{b^2(k)}
     \left[ v_2^2 \rho_+ \eta_1^2 + v_1 v_2 \rho_- \eta_1^2 + 
            v_2^4 \rho_+^3 + 3 v_1 v_2^3 \rho_+^2 \rho_- + 3 v_1^2 v_2^2
                                                          \rho_+ \rho_-^2
            + v_1^3 v_2 \rho_-^3    \right],
                                                               \\ \nonumber
G_2^{\rho_-}&=& \frac{2g}{b(k)}
                [v_1 a_0 \dot{\eta_1} + (v_1^2 - v_2^2) \rho_+ \eta_1^{\prime}
                 - 2 v_1 v_2 \rho_- \eta_1^{\prime}]
                                                         \\ \nonumber
& &+ \frac{2\sqrt{2} \lambda}{b^2(k)} \phi_{sph}
     [v_1^2 \rho_- \eta_1 + v_1 v_2 \rho_+ \eta_1]
   + \frac{2\sqrt{2} g^2}{b^2(k)} \phi_{sph} \rho_- \eta_1,
                                                          \\ \nonumber
G_3^{\rho_-}&=& \frac{g^2}{b^2(k)}
               \Bigg[\rho_- \eta_1^2 + v_1^2 a_0^2 \rho_- + v_1 v_2 a_0^2
                                                            \rho_+
                     + 2 v_1^2 v_2^2 \rho_-^3 + v_1 v_2 (v_1^2 - v_2^2)
                                                                \rho_+^3
                                                         \\ \nonumber
& & \hspace{2.0cm}
                     + (v_1^4 - 4 v_1^2 v_2^2 + v_2^4) \rho_+^2 \rho_-
                     -3 v_1 v_2 (v_1^2 - v_2^2) \rho_+ \rho_-^2  \Bigg]
                                                          \\ \nonumber
& & + \frac{\lambda}{b^2(k)}
        \left[ v_1^2 \rho_- \eta_1^2 + v_1 v_2 \rho_+ \eta_1^2
             + v_1^4 \rho_-^3 + v_1 v_2^3 \rho_+^3 + 3 v_1^2 v_2^2 \rho_+^2
                                                                   \rho_-
             + 3 v_1^3 v_2 \rho_+ \rho_-^2     \right],
                                                          \\ \nonumber
G_2^{\eta_1}&=& -\frac{2g}{b(k)}
               \Bigg[ v_2(\dot{a_0} \rho_+ + a_0 \dot{\rho_+})
                    + v_1(\dot{a_0} \rho_- + a_0 \dot{\rho_-})
                    + 2(v_1 v_1^{\prime} - v_2 v_2^{\prime})\rho_+ \rho_-
                                                          \\ \nonumber      
          & + & (v_1^2 - v_2^2) (\rho_+^{\prime} \rho_- +
                                       \rho_+ \rho_-^{\prime})
                   + v_1^{\prime} v_2 (\rho_+^2 - \rho_-^2)
                   + v_1 v_2^{\prime} (\rho_+^2 - \rho_-^2)
                   + 2 v_1 v_2 (\rho_+ \rho_+^{\prime} - \rho_-
                                                         \rho_-^{\prime}
                                                                  \Bigg]
                                                           \\ \nonumber
& &+ \frac{\sqrt{2} \lambda}{b^2(k)} \phi_{sph}
     (3 \eta_1^2 + v_2^2 \rho_+^2 + v_1^2 \rho_-^2 + 2 v_1 v_2 \rho_+ \rho_-)
+ \frac{\sqrt{2}g^2}{b^2(k)} \phi_{sph} (a_0^2 + \rho_+^2 + \rho_-^2),
                                                            \\ \nonumber
G_3^{\eta_1}&=& \frac{g^2}{b^2(k)} (a_0^2 + \rho_+^2 + \rho_-^2) \eta_1
               + \frac{\lambda}{b^2(k)}
                [\eta_1^3 + (v_2 \rho_+ + v_1 \rho_-)^2 \eta_1].
\end{eqnarray}
Here $z_0 \equiv b(k) \tau$, $z_1 \equiv b(k) x$, and the dot
and the prime denote 
 differentiation with respect to $z_0$ and $z_1$
respectively. Also, the fluctuation
operators $\hat{h}_{a_0}$, $\hat{h}_{\rho_+}$, $\hat{h}_{\rho_-}$, 
and $\hat{h}_{\eta_1}$ are
\begin{eqnarray}
\hat{h}_{a_0}&=&-\frac{\partial^2}{\partial z_1^2} + \frac{2g^2}{b^2(k)}
                                                     \phi_{sph}^2,
                                                          \\   \nonumber
\hat{h}_{\rho_+}&=&-\frac{\partial^2}{\partial z_1^2} + 
                   \frac{1}{b^2(k)}
                  \left[ 2g^2 \phi_{sph}^2 + \lambda (\phi_{sph}^2 - 
                                                      \frac{v^2}{2})
                                          + \lambda \sqrt{f_1}
                                                           \right],
                                                            \\   \nonumber
\hat{h}_{\rho_-}&=&-\frac{\partial^2}{\partial z_1^2} + 
                   \frac{1}{b^2(k)}
                  \left[ 2g^2 \phi_{sph}^2 + \lambda (\phi_{sph}^2 - 
                                                      \frac{v^2}{2})
                                          - \lambda \sqrt{f_1}
                                                           \right],
                                                            \\   \nonumber
\hat{h}_{\eta_1}&=&-\frac{\partial^2}{\partial z_1^2}
                   + \frac{2 \lambda}{b^2(k)} 
                   \left[ 3 \phi_{sph}^2 - \frac{v^2}{2} \right].
\end{eqnarray}
In the following section the resulting fluctuations and the
characteristics of the thermal transitions are analyzed in
detail.

\section{Fluctuation Analysis and Quantum--Classical Transitions} 

Having derived the fluctuation equations, our next aim is to
derive the eigenvalues and then with knowledge of 
the negative mode (as required in the method of Ref. \cite{rana})
 to investigate quantum--classical transitions
and their order.

The lowest few eigenvalues of $\hat{h}_{a_0}$ and $\hat{h}_{\eta_1}$ can
be obtained exactly by
 using Lam$\acute{e}$'s equation  \cite{ars64}. It is easy to 
show that the spectrum of $\hat{h}_{a_0}$ consists of only positive modes
whose explicit forms are not needed here for further study. Also, of the 
lowest eigenstates of $\hat{h}_{\eta_1}$, we need only the 
$2K$--antiperiodic eigenfunctions to recover the proper uncompactified 
limit as shown in Ref.\cite{park00-2}. The lowest two $2K$--antiperiodic
eigenstates
 of $\hat{h}_{\eta_1}$ are summarized in Table I. It may be impossible 
to obtain the higher states analytically at present. Using 
$\int_{-K}^{K} \psi_i^{(\eta_1) \ast} \psi_j^{(\eta_1)} dz_1 = \delta_{ij}$,
one can show (using formulae of Ref. \cite{yellow}) that
 the normalization constant $N_1$ in Table I is given by 
\begin{equation}
N_1 = \sqrt{
            \frac{3k^2}{2[(1 - k^2) K - (1 - 2 k^2) E]}  }.
\end{equation} 

We now consider the eigenstates of $\hat{h}_{\rho_+}$ and $\hat{h}_{\rho_-}$.
In Appendix A we explain how the eigenstates of $\hat{h}_{\rho_+}$
and $\hat{h}_{\rho_-}$ are computed numerically.
 Following the method of Appendix A, 
one can show that the eigenstates of $\hat{h}_{\rho_+}$ also consist of 
only positive modes which we do not need. What we need
(as pointed out earlier), is only
the  negative 
mode of $\hat{h}_{\rho_-}$. If one performs the numerical calculation, 
one finds that $\hat{h}_{\rho_-}$ has two negative modes, one of which
is $2K$--periodic and the other $2K$--antiperiodic. Fig. 1 shows 
the $k$--dependence of the negative eigenvalues for $\theta = 1$. Since the 
$2K$--antiperiodic boundary condition is required for the proper continuum
limit, we have to use the solid line in Fig. 1 as a negative eigenvalue. 
One should note that this negative eigenvalue
 approaches zero in the small $k$ region.
We show in the following that this effect guarantees 
that the instanton--sphaleron transition in the small $k$--region
is different from that
in the large $k$--region. Fig. 2 shows normalized $2K$--antiperiodic 
eigenfunctions for the negative mode of $\hat{h}_{\rho_-}$ at 
($\theta=1$, $k=0.6$) and ($\theta=1$, $k=0.99$).
Their Gaussian shape is indicative of their ground--state
nature (below the zero--eigenvalue of the translational mode).

We let  $\psi_{-1}^{(\rho_-)}$ and $\epsilon_{-1}^{(\rho_-)}$ be
respectively the 
$2K$--antiperiodic eigenfunction and corresponding
eigenvalue for the negative mode. To obtain the criterion 
for the sharp first--order instanton--sphaleron transition we have to 
compute the nonlinear correction to the frequency of the periodic
instanton around the sphaleron. This can be carried out by solving 
Eq.(\ref{expand}) perturbatively. The perturbation procedure is briefly
summarized in Appendix B. The criterion for the first-order transition is 
expressed as an inequality\cite{park99-1,rana}
\begin{equation}
\label{inequality}
\Omega - \Omega_{sph} > 0, 
\end{equation}
where $\Omega$ is the frequency involving the nonlinear correction and 
$\Omega_{sph} \equiv \sqrt{-\epsilon_{-1}^{(\rho_-)}}$.

In Appendix B it is shown
 that the inequality (\ref{inequality}) can be expressed as 
\begin{equation}
\label{criterion}
<\psi_{-1}^{(\rho_-)} \mid D_1(z_1)> \hspace{0.5cm} < \hspace{0.5cm} 0
\end{equation}
where 
\begin{equation}
D_1(z_1) = D_1^{(1)}(z_1) + D_1^{(2)}(z_1) + D_1^{(3)}(z_1).
\end{equation}
Here 
\begin{eqnarray}
D_1^{(1)}(z_1)&=& \frac{2\sqrt{2(1 + k^2)}}{v} 
                  \psi_{-1}^{(\rho_-)}(z_1)
                 \Bigg[ k \left( v_1^2 + \frac{s_1(s_1 + 1)}{2} \right)
                        sn [z_1] g_{\eta_1, 1}(z_1) 
                                                         \nonumber\\
& & \hspace{4.0cm} - 
                        \sqrt{s_1 (s_1 + 1)} v_1 v_2 
                        g_{\eta_1, 1}^{\prime}(z_1)
                                                     \Bigg],
                                                         \nonumber\\
D_1^{(2)}(z_1)&=& \frac{\sqrt{2(1 + k^2)}}{v} 
                  \psi_{-1}^{(\rho_-)}(z_1)
                 \Bigg[ k \left( v_1^2 + \frac{s_1(s_1 + 1)}{2} \right)
                        sn [z_1] g_{\eta_1, 2}(z_1) 
                                                         \nonumber\\
& & \hspace{4.0cm} - 
                        \sqrt{s_1 (s_1 + 1)} v_1 v_2 
                        g_{\eta_1, 2}^{\prime}(z_1)
                                                     \Bigg],
                                                         \nonumber\\
D_1^{(3)}(z_1)&=& \frac{3(1 + k^2)}{4 v^2}
                  [v_1^4 + s_1 (s_1 + 1) v_1^2 v_2^2]
                  \psi_{-1}^{(\rho_-)3}(z_1),
\end{eqnarray}
where $s_1 \equiv \sqrt{\theta^2 + \frac{1}{4}} - \frac{1}{2}$ and
\begin{eqnarray}
\label{gdefine}
g_{\eta_1, 1}(z_1)&=& \hat{h}_{\eta_1}^{-1} \mid q(z_1) >,
                                                         \\   \nonumber
g_{\eta_1, 2}(z_1)&=& (\hat{h}_{\eta_1} + 4 \Omega_{sph}^2)^{-1}
                                            \mid q(z_1) >,
                                                         \\   \nonumber
\mid q(z_1)>&=& -\frac{1}{v} \sqrt{ \frac{1 + k^2}{2}}
                 \bigg[ \theta \left( (v_1 v_2)^{\prime} 
                                      \psi_{-1}^{(\rho_-)2} + 
                                      2 v_1 v_2 \psi_{-1}^{(\rho_-)}
                       \psi_{-1}^{(\rho_-)\prime} \right) \\ \nonumber
          & & + k (v_1^2 + \frac{\theta^2}{2})
                          sn [z_1] \psi_{-1}^{(\rho_-)2}   \bigg].
\end{eqnarray}

It is now necessary to evaluate $g_{\eta_1, 1}$ and $g_{\eta_1, 2}$
explicitly. Although one can calculate $g_{\eta_1, 1}$ exactly by 
following the procedure given in the Appendix of Ref.\cite{park00-2},
this is 
not necessary here. We already know the type of instanton--sphaleron
transition at $k=1$\cite{mat93,park00-1} so that our interest
concerns only 
the domain of small $k$. We can therefore
 adopt the following  
approximate procedure which has been shown to be valid in the small $k$ 
region\cite{park00-2}. Using the completeness relation one can express 
$g_{\eta_1, 1}$ as
\begin{equation}
\label{complete}
g_{\eta_1, 1} = \sum_{n=0}^{\infty}
               \frac{<\psi_n^{(\eta_1)} \mid q >}{\epsilon_n^{(\eta_1)}}
               \mid \psi_n^{(\eta_1)} >.
\end{equation}
Since $\mid q >$ is an odd function, the zero mode of 
$\hat{h}_{\eta_1}$ does not contribute to the r.h.s. of Eq.(\ref{complete}).
Hence the first approximation of $g_{\eta_1, 1}$ is 
\begin{equation}
g_{\eta_1, 1} \approx \frac{<\psi_1^{(\eta_1)} \mid q >}{\epsilon_1^{(\eta_1)}}
               \mid \psi_1^{(\eta_1)} >
\end{equation}
which can be evaluated numerically. In fact, this approximation is valid
when $\mid \psi_1^{(\eta_1)} >$ is an isolated discrete mode and the density of 
higher states is very dilute. Ref.\cite{park00-2} shows these conditions
are fulfilled in the  small $k$--region
 if $\hat{h}_{\eta_1}$ is a Lam$\acute{e}$
operator as is the case here.
 In the same way $g_{\eta_1, 2}$ is approximately
\begin{equation}
g_{\eta_1, 2} \approx \frac{<\psi_1^{(\eta_1)} \mid q >}{3 k^2 
                                                        + 4 \Omega_{sph}^2}
               \mid \psi_1^{(\eta_1)} >.
\end{equation}
The plots of Fig. 3 show the $k$-dependence of 
$$
J_i \equiv <\psi_{-1}^{(\rho_-)} \mid D_1^{(i)} >
$$ 
and of the sum $J_1 + J_2 + J_3$ for $\theta = 1$.
One can see that this sum becomes negative at approximately $k = 0.2$
and therefore satisfies the inequality (\ref{criterion}) and so
(\ref{inequality}) for the existence of a first order transition. 
Thus Fig. 3 demonstrates
 that the sharp first--order
instanton--sphaleron transition occurs at $k < k_c \approx 0.2$ for 
$\theta = 1$. Although the result is not included in this paper, 
we have checked also the $\theta=3$ case and have found a similar behavior:
a sharp transition occurs in the small $k$ region.

\section{Conclusions}

The study of phase transitions is of considerable significance
in many areas of physics, but -- as is also evident from the
above -- this requires highly
 nontrivial efforts, both analytically and numerically.  
In the above we studied a model which permits a considerable fraction of
analytical investigation, but finally requires
also highly nontrivial
 computational work.  The results we presented above answer
the naturally asked question as to
what behavior the Abelian--Higss model would reveal
if the spacial coordinate is compactified on a circle.  We have found that
indeed a change occurs as compared to the
uncompactified case, i.e. in the region
of small elliptic modulus $k$ of the periodic instantons that we used,
which corresponds to circle--circumferences below a
critical size (a specific critical value
was given for appropriate values of other parameters).
Hence, depending on $k$, this model allows both smooth second--order
transitions in the
large $k$ region and sharp first--order transitions in the
small 
$k$ region. These
findings are similar to those
of  $d=4$ SU(2)--Higgs theory in which the type of transition 
depends on the ratio of $M_H$ and $M_W$. 
Thus our findings can be seen as an analogy.  Of course,
our model lacks direct physical application, but this was
also not anticipated. Rather we explored the model also for the other
reasons stated, i.e. as a matter of curiosity as to what type
of thermal behavior will be found once the spatial coordinate
is compactified, and as a further testing ground for methods
of investigation of phase transitions, here in the sense of
transitions from quantum to classical behavior. Such investigations
are usually very complicated and there are few models that 
permit also transparent analytical investigation, at least in part.

\vspace{1cm}

\noindent
{\bf Acknowledgement:} 
This work has been supported in part by
the Korean Research Foundation (Contract Number: KRF-2000-D00073).

\newpage

\centerline{\large\bf Table} 
\vspace{0.2cm}

\begin{table}
\begin{tabular}{|c|c|}\hline\hline
 Eigenvalue of $\hat{h}_{\eta_1}$ &  Eigenfunction of $\hat{h}_{\eta_1}$
                                              \hspace{4.0cm}   \\
                                        \hline 
 $\lambda_0^{(\eta_1)} = 0$ & $\psi_0^{(\eta_1)}(z_1) = N_0 cn[z_1] dn[z_1] $
                                               \hspace{4.0cm}   \\
                                        \hline 
 $\lambda_1^{(\eta_1)} = 3k^2$ & $\psi_1^{(\eta_1)}(z_1) = N_1 sn[z_1] dn[z_1]$
                                              \hspace{4.0cm}     \\
\hline\hline
\end{tabular}
\end{table}

\newpage
\begin{appendix}{\centerline{\bf Appendix A}} 
Here we 
explain how the spectrum of $\hat{h}_{\rho_-}$ is obtained.
The spectrum of $\hat{h}_{\rho_+}$ can be obtained
similarly.
The eigenvalue equation of $\hat{h}_{\rho_-}$ is
\begin{equation}
\label{eigenrho-}
\left[ -\frac{\partial^2}{\partial z_1^2} + f(k, \theta, z_1) \right]
\psi_n^{(\rho_-)} = \zeta \psi_n^{(\rho_-)}
\end{equation}
where
\begin{eqnarray}
f(k, \theta, z_1)&=& (1 + \theta^2) k^2 sn^2 [z_1] - 
                   \sqrt{(1 + 4 \theta^2) \bigg(k^2 sn^2 [z_1] -
 \frac{1 + k^2}{2}\bigg)^2
                         - \theta^2 (1 - k^2)^2 },
                                                   \nonumber\\
\zeta&=& \epsilon^{(\rho_-)} + \frac{1 + k^2}{2}.
\end{eqnarray}
We first choose the $4K$--periodic boundary condition.
 In this case we can use the
Fourier expansions
\begin{equation}
\label{fourier}
f(k, \theta, z_1)= \sum_{n = -\infty}^{\infty}
 a_n e^{i \frac{n\pi}{l} z_1}, \;\;
\psi_n^{(\rho_-)}= \sum_{n = -\infty}^{\infty} b_n e^{i \frac{n\pi}{l} z_1},
\end{equation}
where $l = 2K$ and the coefficients $a_n$ are given by
\begin{equation}
a_n = \frac{1}{2 l} \int_{-l}^{l} f(k, \theta, z_1) e^{-i \frac{n\pi}{l} z_1}.
\end{equation}
Inserting (\ref{fourier}) into (\ref{eigenrho-}) and using
the property of linear independence
of the exponential functions one obtains
\begin{equation}
\sum_{m} \left[ \left( \frac{n \pi}{l} \right)^2 \delta_{mn} + a_{n - m} \right]
    b_m = \zeta b_n.
\end{equation}
Solving this matrix equation numerically, one can evaluate the eigenvalue
$\epsilon_n^{(\rho_-)}$ and eigenfunction $\psi_n^{(\rho_-)}$.
After that we choose only $2K$--antiperiodic eigenfunctions
and determine the corresponding eigenvalues for the proper $k=1$ limit. 
\end{appendix}

\begin{appendix}{\centerline{\bf Appendix B}}
In this appendix we show briefly
 how the inequality (\ref{criterion}) is 
derived for the criterion of the sharp first--order transition by solving 
Eq.(\ref{expand}) perturbatively.
First we choose an {\it ansatz}
\begin{equation}
\label{lowest}
\left( \begin{array}{c}
       a_0 \\ \rho_+ \\ \rho_- \\ \eta_1
       \end{array}                        \right) = \Delta
\left( \begin{array}{c}
       a_{0,0}(z_1) \\ \rho_{+,0}(z_1) \\ \rho_{-,0}(z_1) \\ \eta_{1,0}(z_1)
       \end{array}              \right)  \cos \Omega_{sph} z_0
\end{equation}
where $\Delta$ is a small oscillation amplitude around the sphaleron.
After inserting (\ref{lowest}) into Eq.(\ref{expand}) and neglecting
higher order terms, one obtains
\begin{eqnarray}
\Omega_{sph}&=&\sqrt{-\epsilon_{-1}^{(\rho_-)}},  \\  \nonumber
a_{0,0}&=& 0,  \hspace{2.0cm} \rho_{+,0} = 0  ,    \\  \nonumber
\rho_{-,0}&=& \psi_{-1}^{(\rho_-)},  \hspace{1.5cm} \eta_{1,0} = 0.
\end{eqnarray}
For the next order perturbation we set
\begin{equation}
\label{firstorder}
\left( \begin{array}{c}
       a_0 \\ \rho_+ \\ \rho_- \\ \eta_1
       \end{array}                      \right) = 
\left( \begin{array}{c}
       \Delta^2 a_{0,1}(z_0, z_1) \\ \Delta^2 \rho_{+,1}(z_0, z_1) \\
       \Delta \rho_{-,0}(z_1) \cos \Omega z_0 + \Delta^2 \rho_{-,1}(z_0,z_1) \\
       \Delta^2 \eta_{1,1}(z_0, z_1)
       \end{array}                     \right).
\end{equation}
Inserting Eq.(\ref{firstorder}) into Eq.(\ref{expand}) and considering
only terms up to quadratic order,
 one can show there is no frequency shift to
this order. It is also straightforward to show that $a_{0,1}=0$, 
$\rho_{+,1}=0$, $\rho_{-,1}=0$, and
\begin{equation}
\eta_{1,1} = g_{\eta_1, 1}(z_1) + g_{\eta_1, 2}(z_1) \cos 2 \Omega_{sph} z_0
\end{equation}
where $g_{\eta_1, 1}$ and $g_{\eta_1, 2}$ are given by Eq.(\ref{gdefine}).
For the next order perturbation we set
\begin{equation}
\left( \begin{array}{c}
       a_0 \\ \rho_+ \\ \rho_- \\ \eta_1
       \end{array}            \right) = 
\left( \begin{array}{c}
       \Delta^3 a_{0,2}(z_0, z_1) \\ \Delta^3 \rho_{+,2}(z_0,z_1) \\
       \Delta \rho_{-,0}(z_1) \cos \Omega z_0 + \Delta^3 \rho_{-,2}(z_0, z_1) \\
       \Delta^2 \eta_{1,1}(z_0, z_1) + \Delta^3 \eta_{1,2}(z_0, z_1)
       \end{array}             \right).
\end{equation}
Inserting this  into Eq.(\ref{expand}) and considering
contributions  up to cubic order,
one can show that there is a frequency change in this order given by
\begin{equation}
\Omega_{sph}^2 - \Omega^2 = \Delta^2 
<\rho_{-,0} \mid D_1 >
\end{equation}
which proves Eq.(\ref{criterion}).
\end{appendix}

%\begin{table}
%\begin{tabular}{|c|c|}
% Eigenvalue of $\hat{h}_{\eta_1}$ &  Eigenfunction of $\hat{h}_{\eta_1}$
%                                              \hspace{4.0cm}   \\
%                                        \hline 
% $\lambda_0^{(\eta_1)} = 0$ & $\psi_0^{(\eta_1)}(z_1) = N_0 cn[z_1] dn[z_1] $
%                                               \hspace{4.0cm}   \\
%                                        \hline 
% $\lambda_1^{(\eta_1)} = 3k^2$ & $\psi_1^{(\eta_1)}(z_1) = N_1 sn[z_1] dn[z_1]$
%                                              \hspace{4.0cm}     \\
%\end{tabular}
%\end{table}

\newpage

{\centerline {\large\bf Figure Captions}}

\vspace{0.3cm}

{\centerline {\bf Fig.1}}
\noindent
$k$--dependence of the negative eigenvalues
$\epsilon^{(\rho_-)}$ for $\hat{h}_{\rho_-}$ for
$\theta = 1$. The dotted line and the solid
 line represent the negative eigenvalues
for the $2K$--periodic and $2K$--antiperiodic eigenfunctions respectively.
For the correct $k = 1$ limit we have to choose the solid line as
the negative 
eigenvalue.

\vspace{0.2cm}

{\centerline {\bf Fig.2}}
\noindent
The normalized $2K$--antiperiodic eigenfunctions for the negative
mode of $\hat{h}_{\rho_-}$ for (a) $\theta = 1$, $k = 0.6$, and  
(b) $\theta = 1$, $k = 0.99$.

\vspace{0.2cm}

{\centerline {\bf Fig.3}}
\noindent
$k$--dependence of $J_1$, $J_2$, $J_3$, and $J_1 + J_2 + J_3$ for
$\theta = 1$. This shows that the sharp first--order instanton--sphaleron
transition occurs for $k < k_c \approx 0.2$.

%\begin{figure}
%\caption{$k$--dependence of the negative eigenvalues
%$\epsilon^{(\rho_-)}$ for $\hat{h}_{\rho_-}$ for
%$\theta = 1$. The dotted line and the solid
% line represent the negative eigenvalues
%for the $2K$--periodic and $2K$--antiperiodic eigenfunctions respectively.
%For the correct $k = 1$ limit we have to choose the solid line as
%the negative 
%eigenvalue.}
%\end{figure}

%\begin{figure}
%\caption{The normalized $2K$--antiperiodic eigenfunctions for the negative
%mode of $\hat{h}_{\rho_-}$ for (a) $\theta = 1$, $k = 0.6$, and  
%(b) $\theta = 1$, $k = 0.99$.}
%\end{figure}

%\begin{figure}
%\caption{$k$--dependence of $J_1$, $J_2$, $J_3$, and $J_1 + J_2 + J_3$ for
%$\theta = 1$. This shows that the sharp first--order instanton--sphaleron
%transition occurs for $k < k_c \approx 0.2$.}
%\end{figure}

\newpage
\epsfysize=10cm \epsfbox{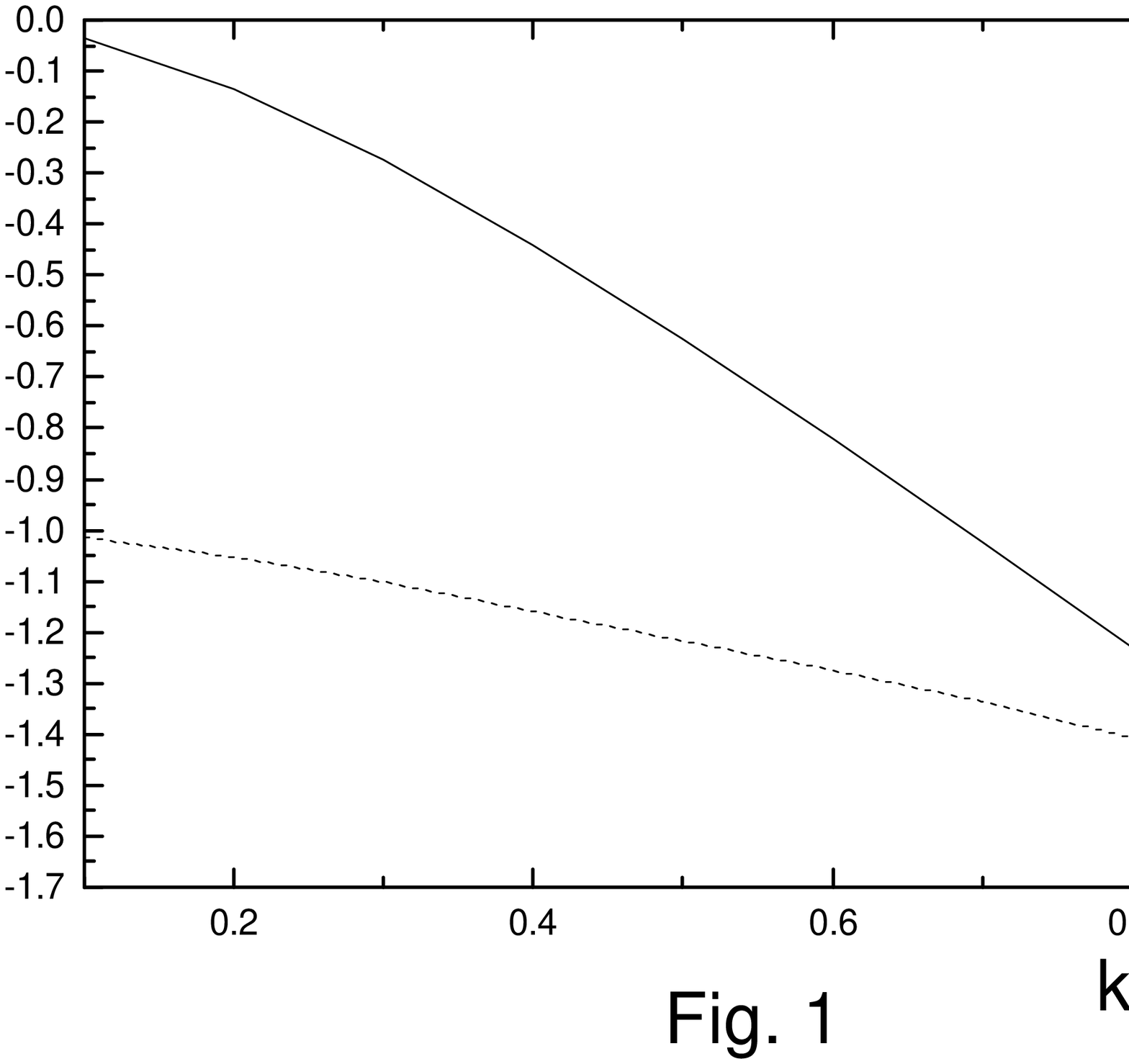}
\newpage
\epsfysize=10cm \epsfbox{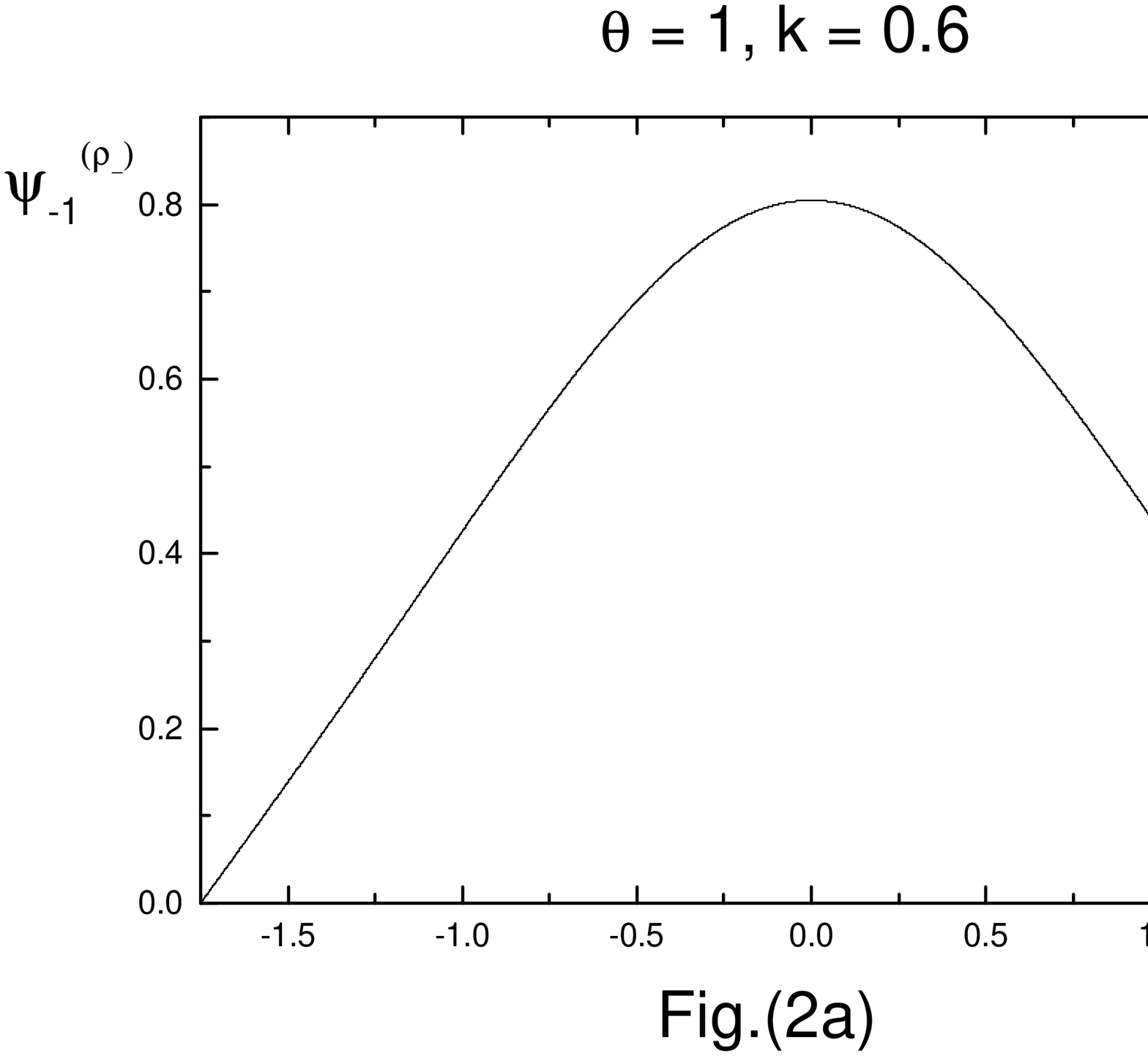}
\epsfysize=10cm \epsfbox{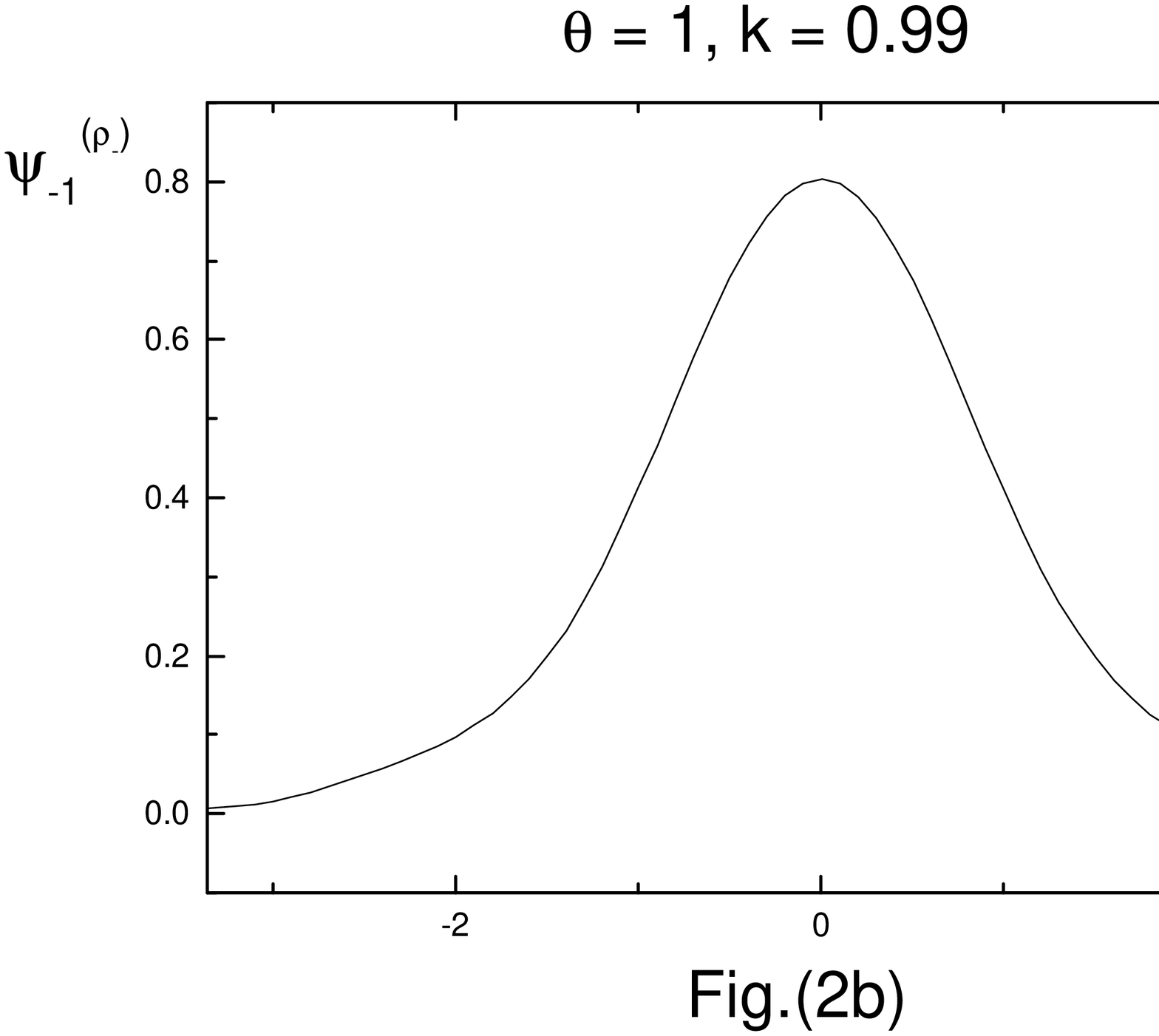}
\newpage
\epsfysize=10cm \epsfbox{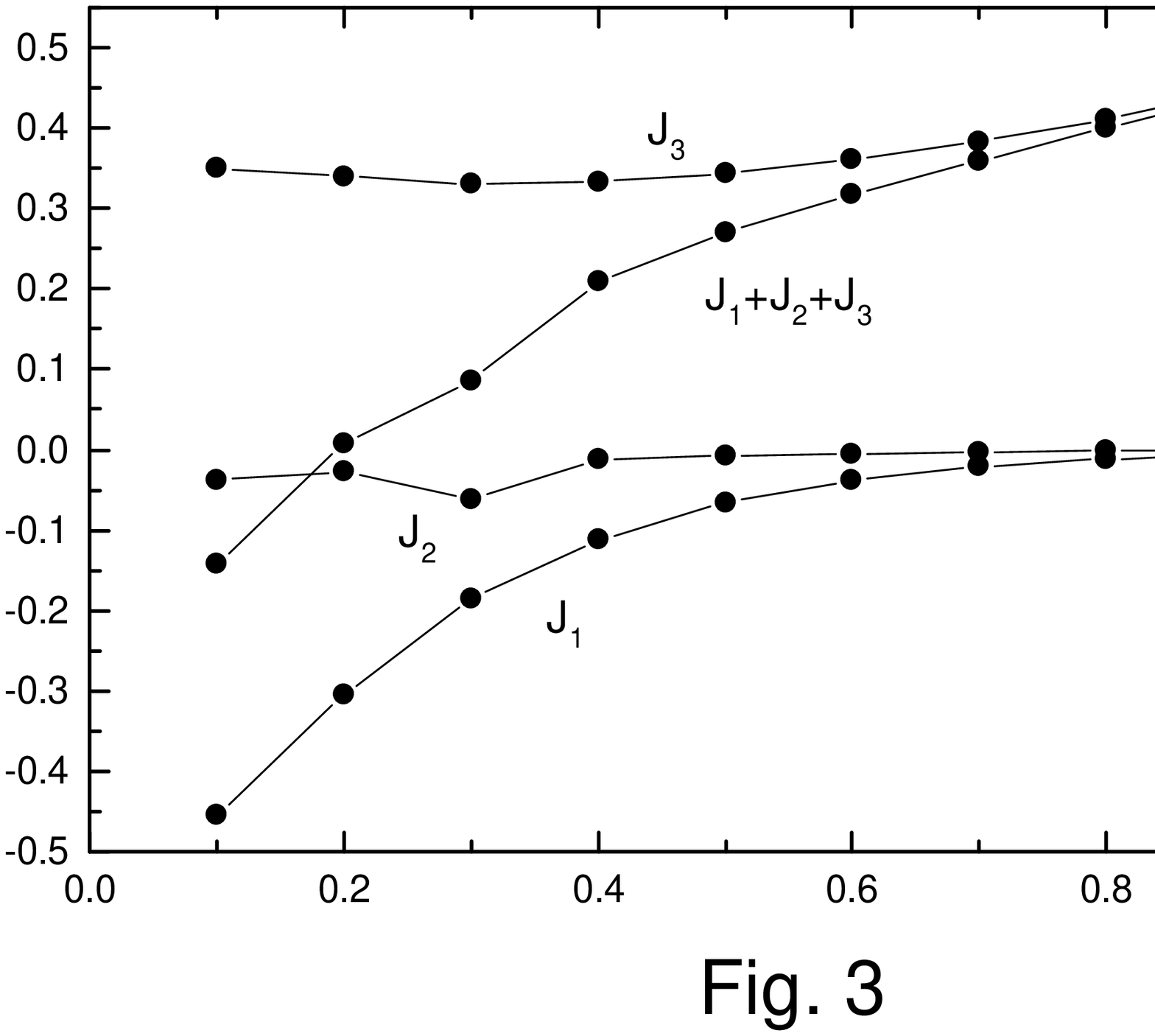}


\begin{thebibliography}{99}
\bibitem{note} By the period of the sphaleron
we  mean the period of the periodic instanton
solution when it approches the sphaleron.
\end{thebibliography}
\end{document}